%% file: main.tex
\documentclass{article}
\pdfoutput=1

\usepackage[utf8]{inputenc}
\usepackage{verbatim}
\usepackage{listings}
\usepackage[letterpaper, margin=1in]{geometry}
\usepackage{braket}
\usepackage{enumitem, kantlipsum}
\usepackage{graphicx}
\usepackage{caption}
\usepackage[dvipsnames,table,xcdraw]{xcolor}
\usepackage{multicol}
\usepackage{mathtools}

\usepackage{csquotes}
\usepackage{amsmath}

\usepackage{subfigure}
\usepackage{url}

\let\OLDthebibliography\thebibliography
\renewcommand\thebibliography[1]{
  \OLDthebibliography{#1}
  \setlength{\parskip}{0pt}
}

\title{An algorithm for DNA read alignment on quantum accelerators}

\author{Aritra Sarkar, Zaid Al-Ars, Carmen G. Almudever, Koen Bertels\\
Department of Quantum \& Computer Engineering\\
Delft University of Technology, The Netherlands}
\date{September 2019}

\begin{document}

\maketitle

\begin{abstract}
    With small-scale quantum processors transitioning from experimental physics labs to industrial products, these processors allow us to efficiently compute important algorithms in various fields. 
    In this paper, we propose a quantum algorithm to address the challenging field of big data processing for genome sequence reconstruction. 
    This research describes an architecture-aware implementation of a quantum algorithm for sub-sequence alignment. 
    A new algorithm named QiBAM (quantum indexed bidirectional associative memory) is proposed, 
    that uses approximate pattern-matching based on Hamming distances.
    QiBAM extends the Grover's search algorithm in two ways to allow for: (1) approximate matches needed for read errors in genomics, and (2) a distributed search for multiple solutions over the quantum encoding of DNA sequences.
    This approach gives a quadratic speedup over the classical algorithm.
    A full implementation of the algorithm is provided and verified using the OpenQL compiler and QX simulator framework. 
    This represents a first exploration towards a full-stack quantum accelerated genome sequencing pipeline design.
    The open-source implementation can be found on \url{https://github.com/prince-ph0en1x/QAGS}.
    
\end{abstract}


\input{section_1.tex} 
\input{section_2.tex} 
\input{section_3.tex} 
\input{section_4.tex} 
\input{section_5.tex} 
\input{section_6.tex} 
\input{section_7.tex} 

\bibliographystyle{unsrt}
\bibliography{ref}

\end{document}

%% file: section_1.tex
\section{Introduction} \label{s1}

The idea of using the fundamental physical building blocks of nature for computation, as proposed by Richard Feynman~\cite{feynman2012there} in 1960s, laid the foundation for the second quantum revolution, focusing on quantum computation instead of the physics of quantum mechanics.
The current development in this field is inspired from two directions - theoretical computer science and computer engineering.


In recent years, there has been a push by the computing industry towards heterogeneous multi-processor systems, where the general-purpose CPU offloads tasks to specialized accelerators, like graphics processing units (GPU), field-programmable gate arrays (FPGA) and digital signal processing units (DSP).
Likewise, we can define a \textit{quantum accelerator} as a classical processor that uses the power of a quantum processor for specific tasks~\cite{riesebos2019quantum}. 
Quantum computing is in essence in-memory-computing, where, the quantum bits stay in principle in the same place and the quantum operations are applied to each qubit.
Thus the logic moves towards the quantum bits in contrary to classical computer accelerators or processors.
In our research, we follow the circuit model for gate-based quantum computing. 
In this design, the classical processor interacts with the quantum processor via commands that specify a sequence of unitary gate operators and receives the collapsed state of the two-level qubit system on measurement. 

In this paper, we use this execution model of the underlying system for application development.
Quantum accelerators allow us to solve a wider complexity class of problems efficiently (the Bounded Quantum Polynomial-time class).
While quantum algorithms are often compared with their classical counterparts in terms of asymptotic complexity, these developments remain fairly independent from the engineering efforts towards manufacturing quantum processing units. 
Current state-of-the-art quantum processors are limited by the number of qubits and coherence time.
There is no clear technological winner among the competing hardware technologies (like superconductors, semiconductors, nitrogen-vacancy centers, ion traps, etc.) in terms of scalability.
The development of quantum algorithms for various use-cases is nevertheless a very active field of research.
These proof-of-concept implementations can be tested for small problem instances on quantum simulators (though we can only go up to around 50 qubits).
For this research, the QX Simulator platform~\cite{khammassi2017qx} with the OpenQL~\cite{openql_2019} language is used.

The most promising candidate applications for quantum acceleration are physical system simulation, cryptography and machine learning.
Our work~\cite{AritraMSc} in this paper falls in the latter category, where the high dimensional state space of the qubits are harnessed to explore/search an optimization landscape faster and better.
Bioinformatics algorithms in use today (especially our focus here, i.e. DNA sequence alignment) rely on heuristic methods to alleviate the huge volume of data.
Even these heuristic approaches take days to run on supercomputing clusters limiting their applicability for more wider use.
The advantage of quantum algorithms discussed in this work is twofold, (a)~they have lower cycle time than corresponding classical algorithms, (b)~the global optima is guaranteed to be sampled with the highest probability instead of sub-optima for current heuristic algorithms.
There is another advantage that puts quantum DNA sequence alignment in the category of near-term applicability.
Application areas like cryptography and physical simulations are highly susceptible to noisy input/computation which makes them highly improbable to achieve good results in the low coherence quantum systems that will be available in the near-term.
The application for DNA alignment tries to search for sub-optima within an acceptable threshold, so an approximate solution is more permissive.

The rest of the paper is organized as follows. 
In the rest of this section, the research problem and the quantum search algorithm are introduced.
Section~\ref{s4} discusses the three existing quantum algorithm designs of associative search, which are used to derive a new search approach in Section~\ref{s5} that incorporate approximate distributed search.
Section~\ref{s6} discusses the results in the context of the application.
Section~\ref{s7} concludes the paper.

%% file: section_2.tex
\subsection{Classical naive DNA sub-sequence alignment} \label{s2}


DNA is a thread-like long polymer made up of nucleic molecules carrying the genetic instructions used in the growth, development, functioning and reproduction of organisms.
The four nucleic molecules, adenine (A), cytosine (C), guanine (G) and thymine (T), represent the building blocks of the DNA. Adenine pairs with thymine and guanine pairs with cytosine, represented by A-T and G-C, which are referred to as base pairs (bp) in the DNA.
The length of genomes varies greatly among organisms, e.g. the human genome is approximately $3.289\times 10^9$ bp long.

\begin{figure}[hb]
    \centering
    \captionsetup{justification=centering}
    \includegraphics[height=2.2in]{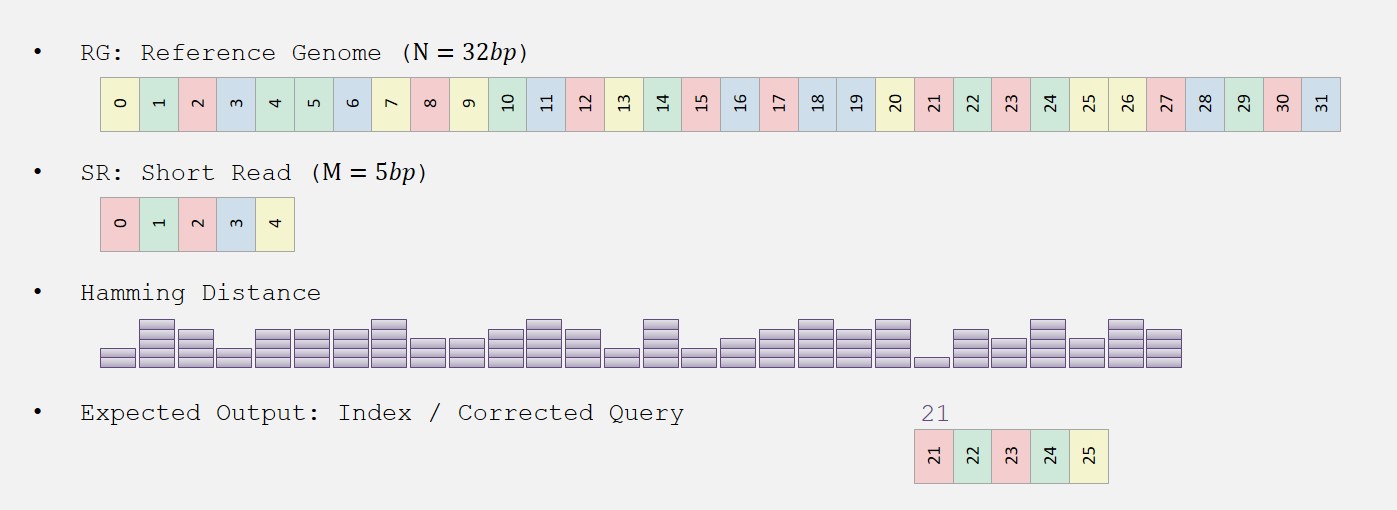}
    \caption{DNA sub-sequence alignment problem}
    \label{f_ssalign}
\end{figure}

The problem we address in this paper is that of DNA sequence alignment.
Sequencers are used to identify the specific sequence of molecules that make up a given DNA sample, which are represented in the form of reads. 
These reads are reconstructed to form the genome and subsequently processed to identify patterns that might indicate specific traits or disease in the organism.
The Broad Institute's GATK DNA processing pipeline~\cite{gatk_2018} is a widely used toolset for this purpose, which includes several processing stages like map-to-reference, duplicate marking and variant calling.
One of the most compute-intensive processing stages is the map-to-reference stage used for aligning the reads for reference-based DNA reconstruction~\cite{houtgast2018hardware}. 
Due to the huge data volume of over-sampled reads, whole-genome sequencing (WGS) of a single human take days on computing clusters, limiting the applicability of WGS in personalized medicine.
This motivates the demand for acceleration using (for example) quantum computation, as even a polynomial speed-up can provide huge benefits on a production environment.

In order to map a sub-sequence of characters (or short read) to a reference sequence, the Levenshtein edit distance is commonly used as a metric for approximate matching of the sub-sequence, spanning the comparison length. The Levenshtein edit distance is upper bounded by the Hamming distance between the two sequences. 
Given the reference sequence $T$ and a short read $P$ of length $N$ and $M$, respectively, the sub-sequence alignment problem is defined as the index $i \in N$ of $T$ where the alignment of $P$ starts, which gives the minimum edit distance.
This is shown in Figure~\ref{f_ssalign}.
The short read is matched for each of the $N-M+1$ starting index in the reference genome.
The alignment algorithm outputs the index of the minimum Hamming distance (here, $21$) and optionally, the nearest match in the reference.

%% file: section_3.tex
\subsection{Quantum search} \label{s3}

In this research, we will be developing on the quantum search algorithm~\cite{grover1996fast}~\cite{grover1997quantum} on an unstructured database, as proposed by Lov Grover.
Grover's search offers a polynomial speedup over a classical linear search (the only search possible for unstructured data).
It is however provably optimal~\cite{zalka1999grover} in query complexity.

\begin{figure}[htb]
    \centering
    \captionsetup{justification=centering}
    \includegraphics[height=0.72in]{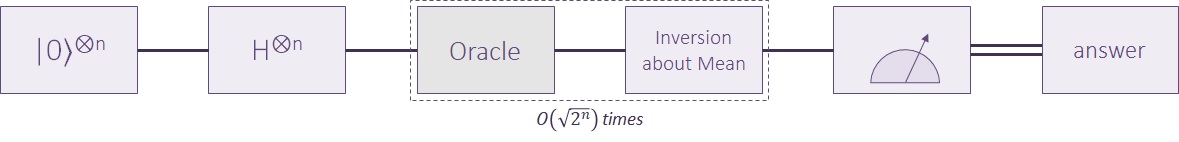}
    \caption{Grover's search steps}
    \label{f_grover}
\end{figure}


Grover's search consists of 3 main steps between state initialization and measurement, as shown in Figure~\ref{f_grover}.
The algorithm creates a uniform superposition of states by applying the Hadamard gate on all qubits in the all-zero state.
The black box Oracle marks the solution state.
Then the amplitude of the solution state is amplified by an inversion-about-mean operation by the Grover gate.
Repeating the last two steps leads to a high probability of measuring the correct solution state.
The search reduces the query complexity to the Oracle by a quadratic factor compared to a linear search.
So, for practical implementation, the gate-complexity of the Oracle also needs to be resolved.

Grover's search was enhanced by subsequent research that will allow us to apply this algorithm in our context.
These improvements are:
\begin{itemize}[nolistsep,noitemsep]
\item Multiple known number of solutions~\cite{boyer1996tight}
\item Arbitrary distribution of initial amplitude~\cite{biham1999grover} 
\item Multiple unknown number of solutions by randomizing iterations over multiple runs~\cite{brassard1998quantum}
\item Multiple unknown number of solutions by a priori counting the number of solutions \cite{john2003sampling}
\end{itemize}
The encoding of an application~\cite{viamontes2005quantum}~\cite{mateus2005quantum} to the Oracle is not explicitly described in these papers, and Grover's original paper assumes the execution of the Oracle in constant time, for an overall polynomial speedup.
However, a complete description of an algorithm in the circuit model needs an explicit representation of the construction of the Oracle with quantum gates as described in this work.

%% file: section_4.tex
\section{Related algorithms} \label{s4}


Associative memory also called content-addressable memory (CAM), is a type of memory organization where instead of the index of the element to be retrieved (similar to a random-access memory, RAM), a partial description of an element is passed as the input query.
The element in the memory with the nearest match to the query is retrieved.

\subsection{Quantum associative memory} \label{s_quam}

\begin{figure}[htb]
    \centering
    \captionsetup{justification=centering}
    \includegraphics[height=0.85in]{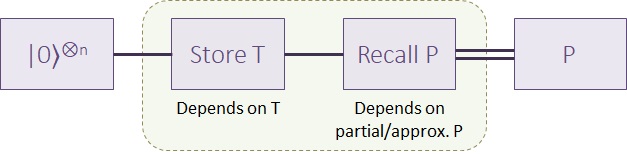}
    \caption{Quantum associative memory algorithm}
    \label{f_quamanatomy}
\end{figure}

The idea of quantum associative memory (QuAM)~\cite{ventura1998quantum, ventura1998artificial, ventura1999initializing, ventura1999quantum, ventura2000quantum} was developed under the umbrella of quantum neural networks (QNN).
Intuitively, the entire parallel search operation is reduced to operations on a superposition of states (memories), resulting in an exponential increase in the capacity of the memory or reduction in the number of comparisons to constant time.
The algorithm consists of two major blocks, a pattern store and a pattern recall as shown in Figure~\ref{f_quamanatomy}.
The pattern store starts from an all-zero initial state.
The information of the reference text string, $T$ is encoded as a superposition of smaller substrings.
This set $T_M$ of length $M$ has substrings $T_M(i)$ (where $i \in \{0\cdots(N-M+1)\}$) made from $T$, each starting from a consecutive index. This is compared with a recall pattern, $P$ representing the query. If an exact match in the stored database is found, it is retrieved as the measurement output. 
However, if a partial or approximate version of $P$ is queried, i.e. some characters of the string are not known exactly, the algorithm returns a random output. Since in practice, $P$ can be inaccurate, our algorithm should be able to retrieve the most similar string from the stored database. 


\subsection{Quantum associative search} \label{s_quamdq}
A major improvement for the quantum associative memory is the use of distributed queries~\cite{ezhov2000quantum}.
Using this concept, the associative memory solves the pattern completion problem; that is, it can restore the full pattern when initially presented with a partial pattern such that the known parts exactly coincide with some part of a valid full pattern.
This allows the associative memory to also retrieve valid memory items when presented with noisy versions of a partial pattern. 
This improvement solves the problem of associative search for which no part of the input stimulus is guaranteed to be noise-free.
It is desirable to retrieve the memory state which is most similar to the given stimulus.
This kind of memory is called a pattern correcting associative memory.

For strings containing only 0s and 1s (binary alphabet), this corresponds to finding the minimum Hamming distance between the query and the memory states.
Amplitudes are distributed in the distributed query such that the maximal value occurs for some definite state $p$ (the provided search query pattern) and the amplitudes of the other states $x$ decrease monotonically with Hamming distance $h(p,x)$.
The binomial distribution matches the required query model.
$p$ is the query center of the binomial distribution.
Let $d=|x|$ be the number of qubits required to store the memory states.
For all $x \in \{0 \cdots (2^d-1)\}$, let $$\ket{b_p^x} = \sqrt{\gamma^{h(p,x)}(1-\gamma)^{d-h(p,x)}}$$
where, $\gamma$ incorporates a metric into the model which tunes the width of the distribution permitting comparison of the similarity of the stimulus and the retrieved memory at a variable scale.
The unitary Oracle transformation can be formed as,
$$O = I_{2^d\times 2^d} - 2\ket{b_p}\bra{b_p}$$
Further modification~\cite{njafa2013quantum} to the model of a distributed query is carried out by merging the concept of the memory state Oracle with the binomial function based Oracle.

\subsection{Quantum indexed memory}
\label{s_qpd}

\begin{figure}[ht]
    \centering
    \captionsetup{justification=centering}
    \includegraphics[height=1.8in]{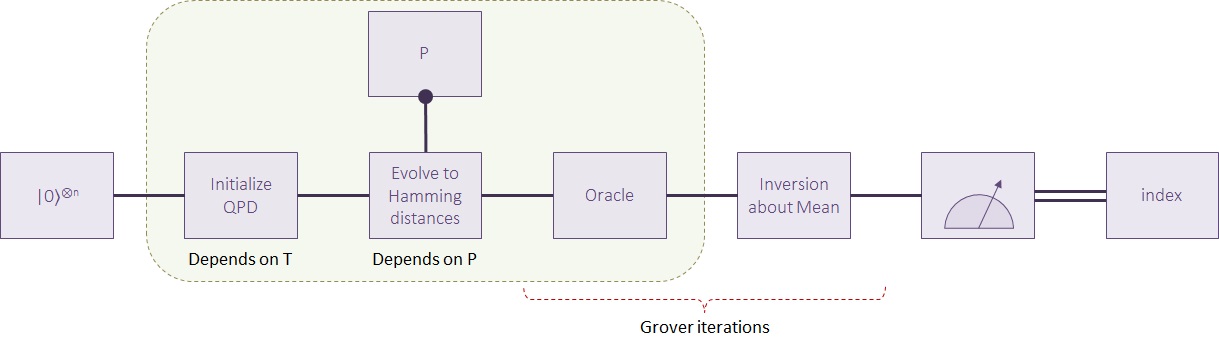}
    \caption{Quantum phone directory algorithm}
    \label{f_qpdanatomy}
\end{figure}

The location in the reference database of an exact or closest match to a query pattern is the alignment index with the minimum Hamming distance between the sequences.
This method~\cite{hollenberg2000fast} was developed for a similar problem of amino-acid sequence matching.
A block diagram of the algorithm is shown in Figure~\ref{f_qpdanatomy}.

The initial state is composed of two quantum registers, the index and the pattern forming the quantum phone directory (QPD) - similar in architecture to a phone directory with name and number.
Essentially, the set of patterns are sorted into an ordered list due to the second register of the database that tags the data.
The initial state is described as,
$$ \ket{\psi_0} = \dfrac{1}{\sqrt{N-M+1}} \sum_{i = 0}^{N-M} \Bigg(\ket{T_M(i)} \otimes \ket{i} \Bigg) $$
where, $T_M(i)$ represents a sub-sequence of the reference $T$ of length $M$ starting at the position $i$.

The next step in the algorithm evolves the data qubits to their Hamming distances with respect to the search pattern $P$.
This operation can be done on the entire superposed state highlighting the parallel transformation power of quantum operators.
A set of CNOT gates with the query pattern $P$ as the control on the data qubits results in the data register evolving to the superposition of Hamming distances between each original data and the query pattern.
The black-box nature of the Oracle function is thus simplified.
For a perfect match, the Oracle now needs to mark the states with the value of $0$, thus making it a fixed function with no dependence on either the reference or the search pattern.
Once the state is amplified according to the modified Grover's algorithm (for an unknown number of solutions), the location of the sequence in the database can be determined by making a measurement on the second part of the entangled register. i.e. the tag qubits.

For approximate matching, the Oracle needs to be modified such that it finds the minimum value of the Hamming distance, instead of an exact $0$.
We propose an algorithm that merges the improved distributed query approach on the indexed quantum data-structure, to retrieve the index of the closest match.

%% file: section_5.tex
\section{Proposed algorithm} \label{s5}

\begin{figure}[ht]
    \centering
    \captionsetup{justification=centering}
    \includegraphics[height=1.95in]{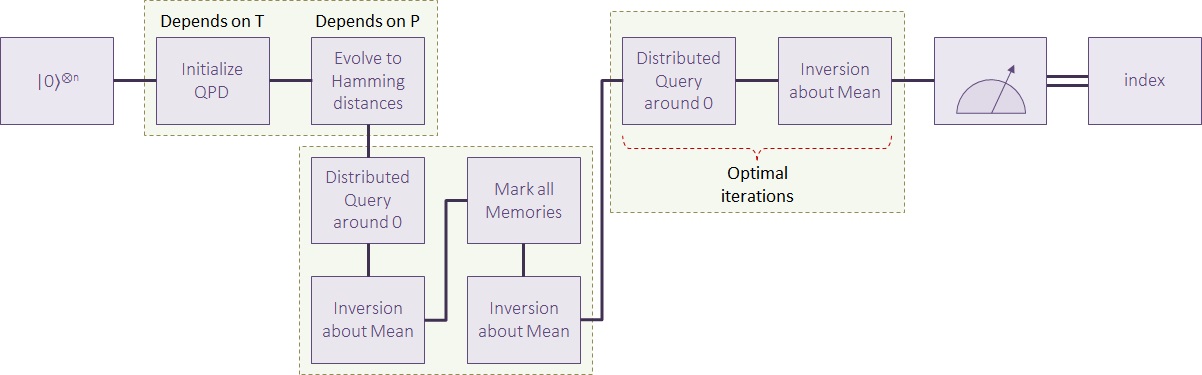}
    \caption{Block diagram of proposed QiBAM algorithm}
    \label{f_qibam}
\end{figure}

The algorithm presented here inherits some of the features from the approaches highlighted in the previous sections.
It is a novel quantum pattern matching algorithm specifically designed for the context of genome sequence reconstruction.
These requirements for the quantum algorithm are:
\begin{itemize}[noitemsep,nolistsep]
    \item In genomic sequences, since reads can contain errors, an approximate matching query is required.
    \item A constant Oracle is useful as compiling the Oracle differently for every read at run-time is tedious.
    \item The associated index in the reference needs to be retrieved, instead of the corrected query.
\end{itemize}
Our proposed algorithm meets these three requirements.
The block diagram for the proposed quantum indexed bidirectional associative memory (QiBAM) algorithm is depicted in Figure~\ref{f_qibam}.

The initialization of the algorithm follows the design as described in Section~\ref{s_qpd}.
The tag qubits encode the pattern index, while the data qubits form the associative memory.
Thus, the pattern store step in the associative memory algorithm (refer Section~\ref{s_quam}) is modelled as a quantum phone directory encoding - which allows the recall of the tagged index corresponding to the query pattern completion/correction. 
Once the data is encoded, the Hamming distance evolution is carried out.
This solves the black-box nature of Grover's marking Oracle.
A distributed query is defined over the associative memory with the query center at zero Hamming distance.

\subsection{Quantum indexed multi-associative memory} \label{s_qimam}

If both the quantum registers are accessible for gate operations, the associative memory can be operated (searched) based on either of the registers thus allowing a bidirectional associative search and retrieval.
We can search with the index (in a RAM mode), or by the data (in a CAM mode).

This idea of associative memory can be generalized to multiple quantum registers holding different attributes of the data that needs to be analyzed.
For example, the quality value of the reads can be stored in register 2, and the chromosome number of the read in register 3, in addition to the index and the pattern.
More complex search queries can be formed based on this entangled quantum database, e.g., search for the index of a specific noisy query pattern among high quality reads in a particular chromosome.

\subsection{Qubit and gate complexity} \label{s_complexity}

The qubit (space) complexity is the aggregate of the qubits used for encoding the data register, tag register and ancilla.
We do not consider overheads for error-correction, mapping, routing or other factors besides the algorithm logic.
For QiBAM, the qubit complexity is the same as the algorithm in Section~\ref{s_qpd}.
Let the number of qubits required for the data and tag registers be $q_d=\left \lceil{log_2(A)}\right \rceil M$ and $q_t=\left \lceil{log_2(N-M)}\right \rceil$ respectively.
The total number of qubits is thus $Q = q_d + q_t + 1$, yielding a typical estimate for the DNA alphabet, the human genome and Illumina reads $A = 4$, $N = 3\times 10^9$ and $M = 50$, as $133$ qubits.

The gate complexity is non-trivial to calculate exactly as it depends on the universal gate set.
Here, the gate set used consists of $\{H,Ry,Rz,C_cX,\}$, where $c=0$ is the X-gate, $c=1$ is the CNOT gate, $c=2$ is the Toffoli gate, and so on.
Higher-order controls can be decomposed with ancilla qubits~\cite{gidney_2018_2}.
The initialization kernel is first decomposed.
First, $q_t$ Hadamard gates are used on the tag qubits to create a superposition of solution states.
Then, conditioned on each tag, the corresponding shifted sub-string of the reference is encoded.
The binary encoding of the tag requires half the controls as inverted, requiring X-gate dressing totalling to $q_t2^{q_t}$.
Using Chargaff's rule, the DNA nucleotides are distributed approximately $1/4$ in each sub-string, requiring $q_d/2$ targets for each tag encoded sub-string.
Thus, the total initialization and Hamming evolution require $q_tH + q_t2^{q_t}C_0X + q_tq_d/2C_{q_t}X$ gates.

The distributed query step depends entirely on the chosen decomposition method for the unitary and the native gate set.
The unitary decomposition method using Quantum Shannon Decomposition (QSD)~\cite{shende2006synthesis} has a complexity of $ 3(4^{n-1}-2^{n-1}) $ where $n$ is the dimension of the unitary.
The mark memory operation would evolve the states in the initial quantum database.
This requires a CPhase over the tag and data qubits for each of the $2^{q_t}$ memories of which $N-M+1$ are memories from the reference genome. 
The data qubits, following Chargaff's rule, would have half the bits as 1, thus using a total of $M$ qubits in average for the compute and uncompute. 
The tag qubits would follow the same behavior as the initialization phase, with average X-dressing of $q_t2^{q_t}C_0X$ gates.
Thus, the total for the marking memory is $(2^{q_t})\{2H+ (M+q_t)C_0X + C_{q_d+q_t-1}X\}$.
The Grover gate is decomposed to $\{2(q_d+q_t)+2\}H + 2(q_d+q_t)C_0X + C_{q_d+q_t-1}X$ gates.

While the exponential reduction in space (qubit) complexity is easy to visualize as proposed in the quantum associative memory architecture, the polynomial speedup is not so pronounced.
This is due to the exponential terms in the worst-case analysis for the QSD and binary encoding.
Many of these gates can be scheduled in parallel in a quantum processor flattening the complexity.
However, since the Oracles are deterministically computable, they can be aggressively optimized by the compiler before run-time.

\subsection{Run-time architecture}

\begin{figure}[h]
    \centering
    \captionsetup{justification=centering}
    \includegraphics[height=1.9in]{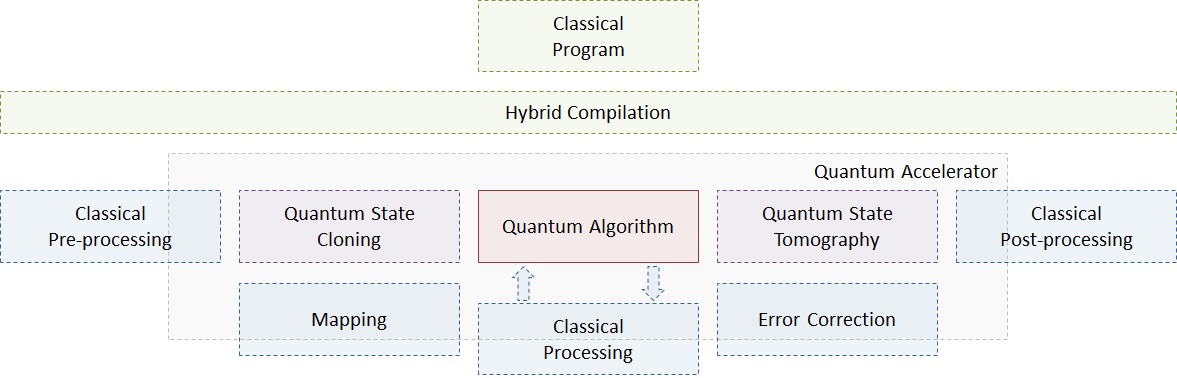}
    \caption{Quantum Algorithm (block with solid outline) and interfacing software architecture}
    \label{f_swdes}
\end{figure}

An isolated discussion of a specific quantum algorithm is not sufficient for near-term implementation.
The quantum algorithm would have interfaces with other software modules running in parallel as shown in Figure~\ref{f_swdes}.

There are 3 factors that contribute to the overall run-time of a general quantum algorithm:
\begin{itemize}[nolistsep,noitemsep]
\item Algorithm: This pertains to the core algorithm running on a simulator, where the internal state vector can be accessed. It refers to the inherent gate complexity of the algorithm and other classical pre/post-processing involved.


\item No-cloning: If the internal state vector cannot be accessed (like in real quantum processors), the experiment needs to be repeated multiple times and the measurement is aggregated.
Most algorithm demands a statistical estimate of the state's probability distribution.
The central tendency of these measurements is the resultant output from the quantum algorithm.

Quantum state tomography is an area of active research.
Advanced methods based on linear inversion, linear regression, maximum likelihood, Bayesian, compressed-sensing and neural networks~\cite{torlai2018neural} exists for estimating the state with fewer tomographic trials.

\item Experimental: For algorithm development (using perfect qubits) and proof-of-concept testing, a simulator platform is preferable, like the QX simulator for this research.
After sufficient confidence in the logic is established, it needs to be ported to an experimental quantum processing unit~\cite{fu2017experimental}.
This adds complexity overhead for topological mapping~\cite{lao2019mapping} and quantum error correction cycles~\cite{varsamopoulos2019decoding}.


\end{itemize}

Thus, every quantum algorithm that depends on a probabilistic result in a noisy environment needs to be repeated, adding a multiplicative factor to the inherent gate complexity.
$$ O(f(\text{experimental}) \times g(\text{no-cloning}) \times h(\text{algorithm})) $$


%% file: section_6.tex
\section{Results on DNA sequences} \label{s6}
The implementation of this algorithm is carried out in OpenQL~\cite{openql_2019} and the QX simulator~\cite{khammassi2017qx}.
The OpenQL framework allows hybrid quantum-classical coding in Python or C++, compiling and optimizing quantum code to produce the intermediate Common QASM (cQASM)~\cite{cqasm} and the compiled Executable QASM (eQASM) for various target platforms (superconducting qubits, spin-qubits, NV-centers, etc.)
The Qxelarator library allows execution of the compiled QASM on the QX binary and receives the measurement outcomes in the high-level OpenQL code encapsulating the quantum architecture (in this case, a simulator) allowing interleaving classical and quantum code blocks in a single program.
QX is a universal quantum computer simulator that takes as input a cQASM file and provides thorough aggressive optimization, high simulation speeds for qubit state evolution.
Our experimental setup (with 28 HT cores, at 2.00 GHz and 384GB memory) can simulate $\approx$35 qubits if the states and operations are non-sparse.
For this algorithm in the development stage, perfect qubits are used without any error model in the QX simulator (like depolarizing model) and the unitary operations have full fidelity.
However, measurement aggregate is used instead of accessing the internal state vector (as allowed for ease of development in most quantum simulators), as this divergence from the ideal distribution has a considerable impact on the algorithm metrics.

\begin{figure}[htb] 
    \centering
    \captionsetup{justification=centering}
    \subfigure[Quantum database for search pattern \emph{CA} and reference string \emph{AATTGTCTAGGCGACC}]
    {%
        \centering
        \includegraphics[width=.46\textwidth]{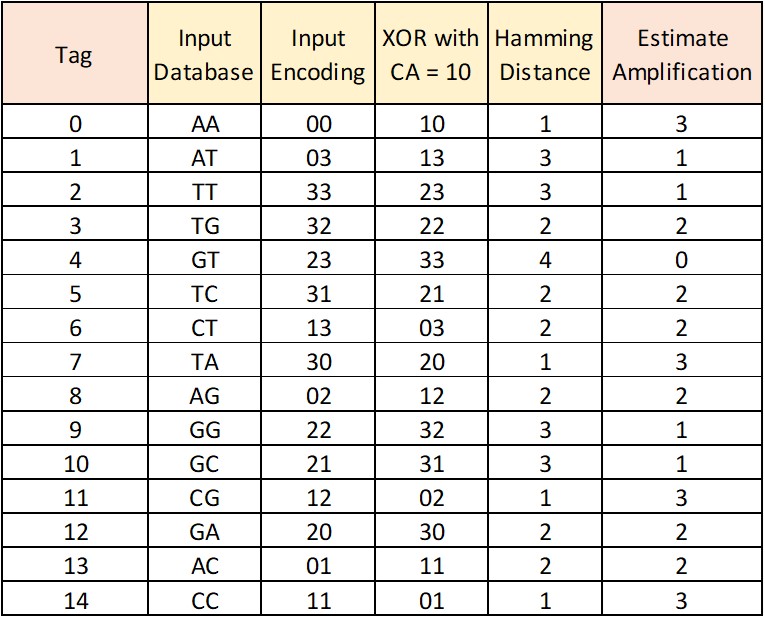}
        \label{f_qibamr1} 
    } 
    \subfigure[Estimate of the solution probability trend]
    {%
        \centering
        \includegraphics[width=.46\textwidth]{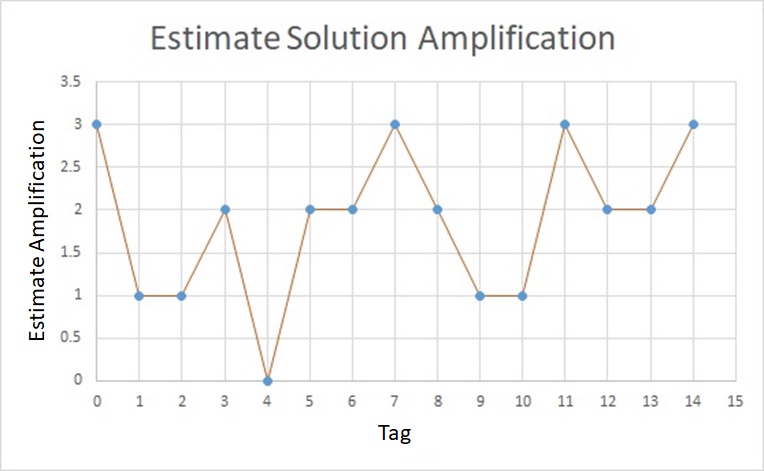}
        \label{f_qibamr2} 
    }
    \caption{Numerical estimate of expected results of a sample run} 
\end{figure}

\begin{figure}[htb] 
    \centering
    \captionsetup{justification=centering}

        \includegraphics[width=.93\textwidth]{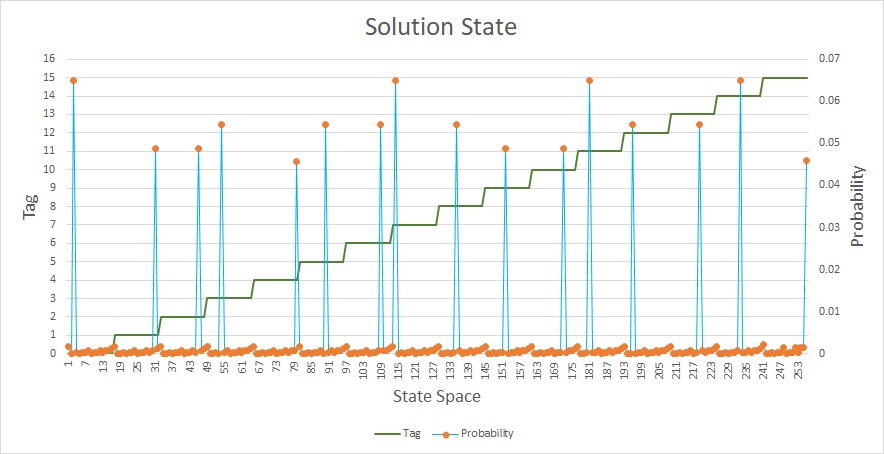}

    \caption{Results of a sample execution of QiBAM on QX Simulator, which matches the results derived analytically in Figure\ref{f_qibamr2}} \label{f_qibamr3}
\end{figure}

In the following example, we will show the results of implementing the QiBAM algorithm on an actual DNA sequence. We show how to search for a pattern of length 2 over the DNA alphabet (\emph{A, C, G, T}). A minimum-length super-string that includes all possible length-2 DNA substrings is \emph{AATTGTCTAGGCGACCA}. This minimal-length super-string helps in verifying the correctness of the quantum algorithm exhaustively.
To test the distributed query capabilities of the algorithm for mismatches in the reference sequence, the last memory, \emph{CA} is not encoded, making the reference genome as \emph{AATTGTCTAGGCGACC}.
This is encoded as the input database shown in Figure~\ref{f_qibamr1}.
Now the search query is chosen as \emph{CA}.
The search pattern conditionally toggles the database to evolve it to the Hamming distance.
Since \emph{CA} is not present in memory, we expect the nearest patterns (approximate matches) to have a higher probability of detection, which are \emph{\{AA, TA, CG, CC\}} (with Hamming distance of 1 in the encoding).
The estimated trend for a higher solution probability should be in line with decreasing Hamming distance, as plotted in Figure~\ref{f_qibamr2}, with the tag on the X-axis and the Estimate Amplification on the Y-axis.

The OpenQL algorithm is executed with the Qxelarator library returning the internal state vector.
The reference sequence and the search query is hardcoded in the Python program for this test but can be streamlined to be directly read from an industry-standard file like the FASTQ format from commercial DNA sequencers.
The result from the run is plotted in Figure~\ref{f_qibamr3}.
The left vertical axis shows the staircase state curve for the tag qubits, while the right vertical axis shows the measurement probability of each individual state.
There are 4 tag qubits and 4 data qubits (2 Radix-4 numbers for a DNA search pattern of length 2).
Thus, the total state space is $2^8 = 256$.
The states with prominent probabilities are the memory states.
The envelope of these states (ignoring the spurious memories) gives the same trend as our estimate in Figure~\ref{f_qibamr2}, verifying the correctness of our implementation.




%% file: section_7.tex
\section{Conclusion} \label{s7}

This research is motivated by the bottleneck of DNA sequence reconstruction in genomics and explores how quantum acceleration can be applied in this domain.
This is the first time a quantum pattern matching algorithm is specifically designed keeping in mind genomic sequences.

The idea of associative memory is extended to an indexed directory of DNA sequences spliced from the reference genome.
In addition to taking into account the DNA alphabet, since reads can contain errors, a distributed query for approximate matching is designed.
This is applied over the superposition of quantum state thereby storing an exponential number of patterns.
A constant Oracle is designed based on minimizing the Hamming distances.
This eliminates the bottleneck of compiling the query differently for every short read at run-time.
The associated index in the reference is retrieved, instead of the corrected query by entangling the index with the sequence database.

This paper also discussed the complexity of the algorithm taking into account system parameters as well.
This algorithm is generalized to a generic quantum data structure for multi-dimensional search.
The algorithm is implemented and verified in the OpenQL environment with the QX Simulator as the backend.

This research is the first exploration~\cite{AritraMSc} towards a roadmap project~\cite{bertels2019quantum} undertaken in the Quantum Computer Architecture lab at the Delft University of Technology, to design a \textit{full-stack quantum accelerator architecture, domain-specific for genome sequencing}.
While the computer application community awaits a large quantum processor capable of real-world problem size execution; our research is carried out on high-performance simulator platforms to test the functional proof-of-concept execution pipeline on small DNA test patterns.
Further research is currently being carried out to adapt the algorithm for near-term quantum computers using parameterized quantum-classical hybrid variational approaches.

%% file: main.bbl
\begin{thebibliography}{10}

\bibitem{feynman2012there}
Richard~P Feynman.
\newblock There's plenty of room at the bottom: An invitation to enter a new
  field of physics.
\newblock In {\em Handbook of Nanoscience, Engineering, and Technology, Third
  Edition}, pages 26--35. CRC Press, 2012.

\bibitem{riesebos2019quantum}
L~Riesebos, X~Fu, AA~Moueddenne, L~Lao, S~Varsamopoulos, I~Ashraf, J~van
  Someren, N~Khammassi, CG~Almudever, and K~Bertels.
\newblock Quantum accelerated computer architectures.
\newblock In {\em 2019 IEEE International Symposium on Circuits and Systems
  (ISCAS)}, pages 1--4. IEEE, 2019.

\bibitem{khammassi2017qx}
Nader Khammassi, I~Ashraf, X~Fu, Carmen~G Almudever, and Koen Bertels.
\newblock Qx: A high-performance quantum computer simulation platform.
\newblock In {\em 2017 Design, Automation \& Test in Europe Conference \&
  Exhibition (DATE)}, pages 464--469. IEEE, 2017.

\bibitem{openql_2019}
Openql framework for high-level quantum programming, 2019.

\bibitem{AritraMSc}
Aritra Sarkar.
\newblock Quantum algorithms for pattern-matching in genomic sequences.
\newblock June 2018.

\bibitem{gatk_2018}
Broad institute gatk best practices pipeline, 2018.

\bibitem{houtgast2018hardware}
Ernst~Joachim Houtgast, Vlad-Mihai Sima, Koen Bertels, and Zaid Al-Ars.
\newblock Hardware acceleration of bwa-mem genomic short read mapping for
  longer read lengths.
\newblock {\em Computational biology and chemistry}, 75:54--64, 2018.

\bibitem{grover1996fast}
Lov~K Grover.
\newblock A fast quantum mechanical algorithm for database search.
\newblock In {\em Proceedings of the twenty-eighth annual ACM symposium on
  Theory of computing}, pages 212--219. ACM, 1996.

\bibitem{grover1997quantum}
Lov~K Grover.
\newblock Quantum mechanics helps in searching for a needle in a haystack.
\newblock {\em Physical review letters}, 79(2):325, 1997.

\bibitem{zalka1999grover}
Christof Zalka.
\newblock Grover’s quantum searching algorithm is optimal.
\newblock {\em Physical Review A}, 60(4):2746, 1999.

\bibitem{boyer1996tight}
Michel Boyer, Gilles Brassard, Peter H{\o}yer, and Alain Tapp.
\newblock Tight bounds on quantum searching.
\newblock {\em arXiv preprint quant-ph/9605034}, 1996.

\bibitem{biham1999grover}
Eli Biham, Ofer Biham, David Biron, Markus Grassl, and Daniel~A Lidar.
\newblock Grover’s quantum search algorithm for an arbitrary initial
  amplitude distribution.
\newblock {\em Physical Review A}, 60(4):2742, 1999.

\bibitem{brassard1998quantum}
Gilles Brassard, Peter H{\o}yer, and Alain Tapp.
\newblock Quantum counting.
\newblock In {\em International Colloquium on Automata, Languages, and
  Programming}, pages 820--831. Springer, 1998.

\bibitem{john2003sampling}
MP~John.
\newblock Sampling with quantum mechanics.
\newblock {\em arXiv preprint quant-ph/0306181}, 2003.

\bibitem{viamontes2005quantum}
George~F Viamontes, Igor~L Markov, and John~P Hayes.
\newblock Is quantum search practical?
\newblock {\em Computing in science \& engineering}, 7(3):62--70, 2005.

\bibitem{mateus2005quantum}
P~Mateus and Y~Omar.
\newblock Quantum pattern matching.
\newblock {\em arXiv preprint quant-ph/0508237}, 2005.

\bibitem{ventura1998quantum}
Dan Ventura and Tony Martinez.
\newblock Quantum associative memory with exponential capacity.
\newblock In {\em Neural Networks Proceedings, 1998. IEEE World Congress on
  Computational Intelligence. The 1998 IEEE International Joint Conference on},
  volume~1, pages 509--513. IEEE, 1998.

\bibitem{ventura1998artificial}
Dan Ventura.
\newblock Artificial associative memory using quantum processes.
\newblock In {\em Proceedings of the International Conference on Computational
  Intelligence and Neuroscience}, volume~2, pages 218--221, 1998.

\bibitem{ventura1999initializing}
Dan Ventura and Tony Martinez.
\newblock Initializing the amplitude distribution of a quantum state.
\newblock {\em Foundations of Physics Letters}, 12(6):547--559, 1999.

\bibitem{ventura1999quantum}
Dan Ventura and Tony Martinez.
\newblock A quantum associative memory based on grover’s algorithm.
\newblock In {\em Artificial Neural Nets and Genetic Algorithms}, pages 22--27.
  Springer, 1999.

\bibitem{ventura2000quantum}
Dan Ventura and Tony Martinez.
\newblock Quantum associative memory.
\newblock {\em Information Sciences}, 124(1-4):273--296, 2000.

\bibitem{ezhov2000quantum}
AA~Ezhov, AV~Nifanova, and Dan Ventura.
\newblock Quantum associative memory with distributed queries.
\newblock {\em Information Sciences}, 128(3-4):271--293, 2000.

\bibitem{njafa2013quantum}
J-P~Tchapet Njafa, SG~Nana Engo, and Paul Woafo.
\newblock Quantum associative memory with improved distributed queries.
\newblock {\em International Journal of Theoretical Physics}, 52(6):1787--1801,
  2013.

\bibitem{hollenberg2000fast}
Lloyd~CL Hollenberg.
\newblock Fast quantum search algorithms in protein sequence comparisons:
  Quantum bioinformatics.
\newblock {\em Physical Review E}, 62(5):7532, 2000.

\bibitem{gidney_2018_2}
Craig Gidney.
\newblock Constructing large controlled nots, 2018.

\bibitem{shende2006synthesis}
Vivek~V Shende, Stephen~S Bullock, and Igor~L Markov.
\newblock Synthesis of quantum-logic circuits.
\newblock {\em IEEE Transactions on Computer-Aided Design of Integrated
  Circuits and Systems}, 25(6):1000--1010, 2006.

\bibitem{torlai2018neural}
Giacomo Torlai, Guglielmo Mazzola, Juan Carrasquilla, Matthias Troyer, Roger
  Melko, and Giuseppe Carleo.
\newblock Neural-network quantum state tomography.
\newblock {\em Nature Physics}, page~1, 2018.

\bibitem{fu2017experimental}
Xiang Fu, MA~Rol, CC~Bultink, J~Van~Someren, Nader Khammassi, Imran Ashraf, RFL
  Vermeulen, JC~De~Sterke, WJ~Vlothuizen, RN~Schouten, et~al.
\newblock An experimental microarchitecture for a superconducting quantum
  processor.
\newblock In {\em Proceedings of the 50th Annual IEEE/ACM International
  Symposium on Microarchitecture}, pages 813--825. ACM, 2017.

\bibitem{lao2019mapping}
Lingling Lao, Daniel~M Manzano, Hans van Someren, Imran Ashraf, and Carmen~G
  Almudever.
\newblock Mapping of quantum circuits onto nisq superconducting processors.
\newblock {\em arXiv preprint arXiv:1908.04226}, 2019.

\bibitem{varsamopoulos2019decoding}
Savvas Varsamopoulos, Koen Bertels, and Carmen~G Almudever.
\newblock Decoding surface code with a distributed neural network based
  decoder.
\newblock {\em arXiv preprint arXiv:1901.10847}, 2019.

\bibitem{cqasm}
cqasm v1.0 towards a common quantum assembly language, 2018.

\bibitem{bertels2019quantum}
K~Bertels, I~Ashraf, R~Nane, X~Fu, L~Riesebos, S~Varsamopoulos, A~Mouedenne,
  H~Van~Someren, A~Sarkar, and N~Khammassi.
\newblock Quantum computer architecture: Towards full-stack quantum
  accelerators.
\newblock {\em arXiv preprint arXiv:1903.09575}, 2019.

\end{thebibliography}
